\documentclass[prl,aps,showpacs]{revtex4}
\usepackage{bm}
\usepackage{amsmath,amssymb}
\usepackage{graphicx}

\begin{document}

\title{Coulombic Energy Transfer and Triple Ionization
in Clusters}
\author{Robin Santra}
\altaffiliation[Present address: ]{JILA, University of Colorado,
Boulder, CO 80309-0440}
\author{Lorenz S. Cederbaum}
\affiliation{Theoretische Chemie, Physikalisch-Chemisches Institut, 
Universit\"{a}t Heidelberg, \\
Im Neuenheimer Feld 229, D-69120 Heidelberg, Germany}
\date{\today}
\begin{abstract}
Using neon and its dimer as a specific example, it is shown that excited Auger decay 
channels that are electronically stable in the isolated monomer can relax in a cluster 
by electron emission. The decay mechanism, leading to the formation of a tricationic cluster, 
is based on an efficient energy-transfer process from the excited, dicationic monomer to a 
neighbor. The decay is ultrafast and expected to be relevant to numerous physical phenomena
involving core holes in clusters and other forms of spatially extended atomic and molecular 
matter.
\end{abstract}
\maketitle

Physical processes accompanying the interaction of X-ray photons with matter are of 
fundamental importance in nature, research, and application. The investigation of clusters 
\cite{SuKo98}, which form a natural bridge between isolated atoms and macroscopic matter, 
offers unique insight into the microscopic workings of cooperative effects taking place
after X-ray absorption. In this letter we report on a novel effect that has not found previous 
consideration. Its consequences are of general relevance for understanding how electron 
correlation in a material drives multiple ionization and fragmentation upon exposure to high-energy 
radiation.

R\"{u}hl {\em et al.} carried out a series of experiments in which they created core 
holes in the 2p shell of argon clusters of different sizes \cite{RuSc91,RuHe94}. They 
investigated the level of fragmentation of the clusters and concluded that, after 
the monomer with the initial core hole becomes dicationic upon Auger electron emission,
one of the two positive charges is transferred to a neighboring atom---a relatively
slow process taking place on a timescale of picoseconds, since energy conservation
requires nuclear motion to set in. In small clusters, the consequence is fragmentation 
through Coulomb explosion. Dicationic clusters exceeding a critical size can be
stable. It must be pointed out, however, that analyzing data from experiments on a 
neutral cluster beam is very difficult because the distribution of cluster sizes in
a beam is rather broad \cite{SuKo98}. Only the average size can be adjusted. It is,
therefore, not trivial to correlate mass-spectroscopically observed fragments with a 
parent cluster of fixed size. In fact, multicoincidence mass spectra published later
by R\"{u}hl and co-workers clearly demonstrated the relevance of {\em triple ionization} 
processes in core-excited clusters \cite{RuKn96} but did not reveal the underlying
physics.

We have found an efficient, new mechanism for triple ionization of clusters, following 
the absorption of a single X-ray photon. Our argument relies on the fact that many of 
the dications produced in Auger decay are not in their electronic ground state but are 
highly excited. In order to clarify our point, we pick a specific example and restrict 
the discussion to neon and its dimer. It will become clear in the course of our 
presentation how most conclusions can be extended to larger neon aggregates as well as 
to clusters consisting of monomers other than neon. This underscores the general nature
of the new process.

\begin{table}[h]
\caption[]{States of Ne$^{++}$, populated by Auger decay of Ne$^+$(1s$^{-1}$)
\cite{Kell75}. The energies are given relative to the energy of neutral Ne
in its ground state (DIP: double ionization potential; AI: Auger intensity).}
\label{tab1}
\begin{tabular}{c|c|c}
 & DIP [eV] & AI [arb. units] \\
\hline
2p$^{-1}$2p$^{-1}$ $^1$D & 65.7  & 10.1 \\
2p$^{-1}$2p$^{-1}$ $^1$S & 68.9  & 1.6  \\
2s$^{-1}$2p$^{-1}$ $^3$P & 87.8  & 1.0  \\
2s$^{-1}$2p$^{-1}$ $^1$P & 98.5  & 2.8  \\
2s$^{-1}$2s$^{-1}$ $^1$S & 122.1 & 1.0  \\
\end{tabular}
\end{table}

There are three conceivable routes to producing Ne$^{3+}$ by removal of a core
electron from a neon atom. First, in the shake-off band of the photoelectron
spectrum, the core hole is accompanied by a valence hole plus a free electron, a 
process commonly interpreted as a consequence of the contraction of the atomic 
shells in the presence of the core hole \cite{KrCa64}. Auger decay of the 
core-excited dication yields a trication. Second, the core-hole main state can 
undergo double Auger decay \cite{CaKr65}, thereby emitting two correlated electrons.
At a photon energy of 1.5 keV, for instance, these two processes are responsible
for the triple ionization of about $21\%$ of all neon atoms that have absorbed an
X-ray photon \cite{LiVa71} (see also \cite{SaSu92}).

\begin{figure}
\includegraphics[width=5.5cm,origin=c,angle=-90]{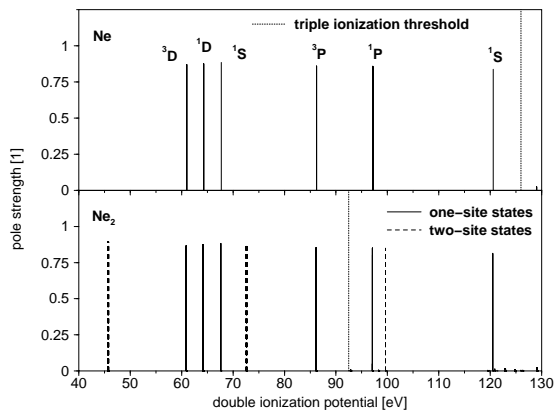}
\caption[]{Double ionization potentials of Ne and Ne$_2$, computed by means
of the ADC(2) approximation of the two-particle Green's function \cite{ScBa84,GoCe96}.
The respective triple ionization thresholds are indicated with dotted lines.}
\label{fig1}
\end{figure}

The third process one might think of is autoionization of some of the Auger decay 
channels themselves. These channels are shown in Table~\ref{tab1}. The data are taken from 
Ref.~\cite{Kell75}. We have converted the Auger energies to double ionization
potentials by making use of the binding energy, $870.2$ eV, of a neon 1s electron
\cite{Sieg69}. The Auger transition to the $^3$D ground state of Ne$^{++}$ is 
suppressed for symmetry reasons. Note that almost $30\%$ of all dications produced 
in Auger decay of Ne$^+$(1s$^{-1}$) are highly excited, possessing at least one hole 
in the inner-valence (2s) shell. Yet, not one of these can relax by electron emission, 
since the triple ionization threshold, $126.0$ eV \cite{SaAn88}, of a single, isolated 
neon atom is energetically too high.

The easiest way to lower the triple ionization threshold---without significantly
modifying the core ionization and the subsequent Auger process---is to add another
neon atom, i.e., to consider Ne$_2$ (or larger neon clusters). Indeed, owing to the 
weak van der Waals interaction between the two neon atoms, both removal of a 1s electron
as well as Auger decay of the resulting core hole are essentially of atomic nature. As
illustrated by the double ionization spectra of Ne and Ne$_2$ in Fig.~\ref{fig1}, the 
relevant double ionization channels in a neon atom are hardly affected by the presence
of an adjacent Ne. The interatomic distance in Ne$_2$ was taken to be $3.2$ \AA, 
corresponding to the equilibrium geometry in the ground state of the van der Waals 
system \cite{MoWi99}. The two-site states of (Ne$_2$)$^{++}$, which are characterized 
by one positive charge on each atom, are not expected to become populated under the 
circumstances considered here. 

We calculated the double ionization spectra using the
{\em algebraic diagrammatic construction} (ADC) scheme for the two-particle
propagator \cite{ScBa84,GoCe96}. ADC({\em n}) represents a sophisticated 
perturbation-theoretical approximation of a many-body Green's function, which is complete 
up to {\em n}th order and includes in addition infinite summations over certain classes of 
diagrams contributing to the expansion. Here, we employed ADC(2); all orbital energies
and molecular integrals needed were taken from closed-shell Hartree-Fock calculations.
On the center of each neon atom we placed the Gaussian basis set d-aug-cc-pVTZ
\cite{WoDu94}. Core polarization effects were investigated and found to be negligible. 
The calculated double ionization energies for the Auger channels of the single atom 
are in good agreement with Table~\ref{tab1}. The pole strengths shown in 
Fig.~\ref{fig1} are related to the residues of the two-particle propagator at its
poles. Values smaller than one indicate for the main states an admixture 
of three-hole one-particle excitations to the two-hole configurations.

The triple ionization threshold of Ne$_2$ is obviously lower than that of the 
isolated atom, because in the dimer the three positive charges do not have to 
be localized to a single atomic site; however, it remains to be proven that this
threshold is lower than any of the inner-valence excited one-site states of 
(Ne$_2$)$^{++}$. To that end we performed multireference configuration interaction
calculations with a program written by Hanrath and Engels \cite{HaEn97}. The basis
set used was again d-aug-cc-pVTZ. For (Ne$_2$)$^{3+}$, the reference space was chosen
to consist of all three-hole configurations with respect to the Hartree-Fock ground
state of neutral neon dimer. We first represented the electronic Hamiltonian in the 
reference space and computed the energies of the lowest doublet and quartet states,
which correlate with Ne$^+$(2p$^{-1}$)~$^2$P + Ne$^{++}$(2p$^{-1}$2p$^{-1}$)~$^3$D.
Then we augmented the many-electron basis by including additionally all single 
excitations of the reference space, i.e., all four-hole one-particle configurations, 
and, finally, all double excitations (five-hole two-particle configurations).
The results are summarized in Table~\ref{tab2}. On the energy scale shown, the 
lowest doublet state and the lowest quartet state of (Ne$_2$)$^{3+}$ are virtually 
degenerate.

\begin{table}
\caption[]{The energies of select states of (Ne$_2$)$^{3+}$ and (Ne$_2$)$^{++}$,
calculated with the method of multireference configuration interaction. The respective
reference spaces, denoted 0, are described in the text; 1 symbolizes single, 2 double
excitations. For (Ne$_2$)$^{3+}$, the energy of its lowest state (doublet and quartet)
is shown. The energy of (Ne$_2$)$^{++}$ refers to its one-site singlet state of
2s$^{-1}$2p$^{-1}$ character.
}
\label{tab2}
\begin{tabular}{c|c|c}
& (Ne$_2$)$^{3+}$ & (Ne$_2$)$^{++}$ \\
configuration space & energy [a.u.] & energy [a.u.] \\
\hline
0 & -253.31 & -253.09 \\
0 + 1 & -253.89 & -253.58 \\
0 + 1 + 2 & -254.12 & -253.89 \\
\end{tabular}
\end{table}

For the dicationic states of neon dimer we pursued a similar strategy. In this case
the reference space comprised all two-hole configurations. We found that the one-site
states associated with the atomic Ne$^{++}$(2s$^{-1}$2p$^{-1}$)~$^1$P and the 
Ne$^{++}$(2s$^{-1}$2s$^{-1}$)~$^1$S Auger channels are indeed above the triple ionization
threshold, as demonstrated by the corresponding entry in Table~\ref{tab2}. The total energies
of (Ne$_2$)$^{++}$ and (Ne$_2$)$^{3+}$ converge only slowly as a function of configuration 
space, but the energy difference between the two species is obviously positive: In all 
configuration spaces studied, the inner-valence excited one-site state of (Ne$_2$)$^{++}$ is 
systematically higher by at least 6 eV than the triple ionization threshold of neon dimer. We 
conclude this state, which is rather efficiently populated in the Auger decay of neon core 
holes (cf. Table~\ref{tab1}), is electronically unstable. 

The triple ionization threshold of Ne$_2$ can be estimated to be---at most---92.5 eV.
Assuming the Coulomb interaction between one Ne$^+$ and one Ne$^{++}$, each being in its
respective ground state, to be that of two point-like charges separated by 3.2 \AA,
we obtain approximately the same value. This indicates the pronounced charge localization 
in the tricationic dimer: One positive charge is on one atom, two positive charges are on 
the other. We exploit this and the fact that one-site states in (Ne$_2$)$^{++}$ are
practically indistinguishable from purely atomic Auger channels (see Fig.~\ref{fig1}) 
for a simple analysis of the mechanism underlying the electronic decay of singlet
(Ne$_2$)$^{++}$.

That atom carrying the outer and the inner valence hole is labeled $e$; the atom remaining
in its neutral ground state is labeled $g$. Formally, the initial singlet state, 
$\left|I\right\rangle$, of the dication can therefore be written as 
\begin{equation}
\label{eq1}
\left|I\right\rangle = \frac{1}{\sqrt{2}}\left\{
\hat{c}_{{\mathrm{2p}},+}^{(e)}\hat{c}_{{\mathrm{2s}},-}^{(e)} -
\hat{c}_{{\mathrm{2p}},-}^{(e)}\hat{c}_{{\mathrm{2s}},+}^{(e)}\right\}
\left|\Phi_0\right\rangle,
\end{equation}
where $\left|\Phi_0\right\rangle$ denotes the Hartree-Fock ground state of neutral Ne$_2$.
$\hat{c}_{{\mathrm{2p}},+}^{(e)}$ is an annihilation operator that removes an electron with 
spin quantum number $m_{{\mathrm{S}}}=+1/2$ from a 2p orbital on atom $e$.
As we have shown, there is at least one tricationic state in the dimer---not in the 
isolated atom---that is lower in energy than $\left|I\right\rangle$. The three open shells
in Ne$^+$(2p$^{-1}$)~$^2$P + Ne$^{++}$(2p$^{-1}$2p$^{-1}$)~$^3$D give rise to two doublets.
Hence, the final state, $\left|F\right\rangle$, including the emitted decay electron, is a
linear combination of two singlet states:
\begin{eqnarray}
\label{eq2}
\left|F_1\right\rangle & = & \frac{1}{2\sqrt{3}}\left\{
\hat{c}_{{\bm k},+}^{\dag}\left(
\hat{c}_{{\mathrm{2p}},-}^{(e)}\hat{c}_{{\mathrm{2p'}},+}^{(e)}
\hat{c}_{{\mathrm{2p}},+}^{(g)}+\hat{c}_{{\mathrm{2p}},+}^{(e)}
\hat{c}_{{\mathrm{2p'}},-}^{(e)}\hat{c}_{{\mathrm{2p}},+}^{(g)}
-2\hat{c}_{{\mathrm{2p}},+}^{(e)}\hat{c}_{{\mathrm{2p'}},+}^{(e)}
\hat{c}_{{\mathrm{2p}},-}^{(g)}\right)\right. \nonumber \\
& & \left.-\hat{c}_{{\bm k},-}^{\dag}\left(
\hat{c}_{{\mathrm{2p}},+}^{(e)}\hat{c}_{{\mathrm{2p'}},-}^{(e)}
\hat{c}_{{\mathrm{2p}},-}^{(g)}+\hat{c}_{{\mathrm{2p}},-}^{(e)}
\hat{c}_{{\mathrm{2p'}},+}^{(e)}\hat{c}_{{\mathrm{2p}},-}^{(g)}
-2\hat{c}_{{\mathrm{2p}},-}^{(e)}\hat{c}_{{\mathrm{2p'}},-}^{(e)}
\hat{c}_{{\mathrm{2p}},+}^{(g)}\right)
\right\}
\left|\Phi_0\right\rangle
\end{eqnarray}
and
\begin{eqnarray}
\label{eq3}
\left|F_2\right\rangle & = & \frac{1}{2}\left\{
\hat{c}_{{\bm k},+}^{\dag}\left(
\hat{c}_{{\mathrm{2p}},-}^{(e)}\hat{c}_{{\mathrm{2p'}},+}^{(e)}
\hat{c}_{{\mathrm{2p}},+}^{(g)}-\hat{c}_{{\mathrm{2p}},+}^{(e)}
\hat{c}_{{\mathrm{2p'}},-}^{(e)}\hat{c}_{{\mathrm{2p}},+}^{(g)}
\right)\right. \\
& & \left.-\hat{c}_{{\bm k},-}^{\dag}\left(
\hat{c}_{{\mathrm{2p}},+}^{(e)}\hat{c}_{{\mathrm{2p'}},-}^{(e)}
\hat{c}_{{\mathrm{2p}},-}^{(g)}-\hat{c}_{{\mathrm{2p}},-}^{(e)}
\hat{c}_{{\mathrm{2p'}},+}^{(e)}\hat{c}_{{\mathrm{2p}},-}^{(g)}
\right)
\right\}
\left|\Phi_0\right\rangle. \nonumber
\end{eqnarray}
The operator $\hat{c}_{{\bm k},+}^{\dag}$ creates a free electron with wavevector
${\bm k}$ and $m_{{\mathrm{S}}}=+1/2$. $\mathrm{2p'}$ is a second 2p orbital
on $e$. The expansion coefficients of $\left|F_1\right\rangle$ and $\left|F_2\right\rangle$
in $\left|F\right\rangle$ are of comparable magnitude. Other contributions resulting
from configuration interaction are less important and disregarded for simplicity.

The transition from the initial to the final state is mediated by electron correlation,
i.e., by the influence of the two-particle Coulomb interaction $\hat{V}$. According
to Wigner and Weisskopf's famous theory of decaying states \cite{WeWi30}, the associated 
decay width is related to the modulus squared of the coupling matrix element 
$\left\langle F \right|\hat{V}\left|I\right\rangle$. For the two components of 
$\left|F\right\rangle$ we find:
\begin{equation}
\label{eq4}
\left\langle F_1 \right|\hat{V}\left|I\right\rangle =
-\sqrt{\frac{3}{2}}\int \int \varphi_{{\bm k}}^{\ast}({\bm x}_1)
{\varphi_{{\mathrm{2s}}}^{(e)}}^{\ast}({\bm x}_2)
\frac{e^2}{|{\bm x}_1 - {\bm x}_2|}
\varphi_{{\mathrm{2p'}}}^{(e)}({\bm x}_1) \varphi_{{\mathrm{2p}}}^{(g)}({\bm x}_2) 
{\mathrm{d}}^3x_1 
{\mathrm{d}}^3x_2,
\end{equation}
\begin{eqnarray}
\label{eq5}
\left\langle F_2 \right|\hat{V}\left|I\right\rangle & = & 
\sqrt{2}\int \int \varphi_{{\bm k}}^{\ast}({\bm x}_1)
{\varphi_{{\mathrm{2s}}}^{(e)}}^{\ast}({\bm x}_2)
\frac{e^2}{|{\bm x}_1 - {\bm x}_2|}
\varphi_{{\mathrm{2p}}}^{(g)}({\bm x}_1)\varphi_{{\mathrm{2p'}}}^{(e)}({\bm x}_2)
{\mathrm{d}}^3x_1
{\mathrm{d}}^3x_2 \nonumber \\
& & -\frac{1}{\sqrt{2}}\int \int \varphi_{{\bm k}}^{\ast}({\bm x}_1)
{\varphi_{{\mathrm{2s}}}^{(e)}}^{\ast}({\bm x}_2)
\frac{e^2}{|{\bm x}_1 - {\bm x}_2|}
\varphi_{{\mathrm{2p'}}}^{(e)}({\bm x}_1) \varphi_{{\mathrm{2p}}}^{(g)}({\bm x}_2)
{\mathrm{d}}^3x_1
{\mathrm{d}}^3x_2.
\end{eqnarray}
Eq.~(\ref{eq4}) and the second term of Eq.~(\ref{eq5}) 
describe an electron-transfer process from the neutral atom to the inner-valence hole 
of Ne$^{++}$. The electron-transfer integral is {\em very} small because of the negligible 
overlap between the spatially compact 2s orbital on the dication and occupied orbitals on 
the adjacent atom.

The decay mechanism is best represented by the first term of Eq.~(\ref{eq5}). It allows
us to formulate the following physical picture. An electron from a 2p orbital on Ne$^{++}$
drops into the inner-valence hole. The released energy is transferred, via virtual-photon
exchange, to the neon atom. As a consequence, an electron is ejected with a kinetic energy 
of about $6$ eV, which is energetically feasible owing to the spatial separation of the 
resulting hole from the two positive charges on the atomic dication. The tricationic dimer 
eventually undergoes Coulomb explosion. 

The mechanism described is analogous to the one discovered in singly ionized, inner-valence
excited hydrogen-bonded \cite{CeZo97} and van der Waals clusters \cite{SaZo00}. This mechanism 
is called Interatomic Coulombic Decay (ICD), which we now have shown to be active in a previously
unexpected context. A detailed Wigner-Weisskopf analysis of ICD in monocationic clusters is 
presented in Ref.~\cite{SaZo01}. The coupling matrix element 
$\left\langle F \right|\hat{V}\left|I\right\rangle$ derived there is similar to the one given
above, if we assume the wavefunction of the ICD electron is comparable in both cases. In fact,
the kinetic energy of the ICD electron here is also just a few electronvolts, corresponding
to a de Broglie wavelength of the order of 1 nm, so that cancellation effects in the coupling 
matrix element, which might arise due to the oscillatory behavior of the free electron's 
wavefunction, do not play a role. On this basis, we expect decay lifetimes of inner-valence 
excited, dicationic clusters of the order of 100 fs or shorter. 

For the purpose of demonstrating this to be correct, we utilized a technique called CAP/CI 
\cite{SoRi98}, i.e., the combination of multireference configuration interaction with a 
complex absorbing potential (CAP) \cite{RiMe93}. A review of non-Hermitian electronic 
theory can be found in Ref.~\cite{SaCe02}. CAP/CI allows the direct {\em ab initio} calculation 
of the Siegert energy of a decaying state, $E_{\mathrm{res}}=E_{R} - {\mathrm{i}}\Gamma/2$,
where $E_{R}$ is the resonance position and $\Gamma$ the decay width. The computations are 
rather challenging and expensive, so we reduced the compact part of the Gaussian basis set to 
d-aug-cc-pVDZ \cite{WoDu94}, but added a set of diffuse basis functions: three s, p, and d 
functions, respectively. These are needed to improve the representation of the outgoing ICD 
electron being absorbed by the CAP. Within a configuration space consisting of all two-hole and 
three-hole one-particle configurations, we calculated a decay lifetime of about 80 fs for each 
of the six one-site, inner-valence excited $^1$P states of (Ne$_2$)$^{++}$.

Using a model system---the neon dimer---we have shown that the possibility of charge separation
in a cluster can give rise to a new, energy-transfer based decay mechanism accessible to a 
number of Auger decay channels. In our specific example, more than $20\%$ of all dications produced 
in the Auger process will undergo ICD and form trications. There are no reasons to suppose our 
simple and general arguments be restricted to neon clusters. We have found first indications 
that a similar effect exists in water clusters as well. The analogy to ICD in singly ionized systems 
implies an interesting dependence of the ICD lifetime on the number of nearest neighbors \cite{SaZo01}: 
Since each neighboring monomer contributes decay channels, we expect the lifetime to drop significantly 
as neighbors are added to an inner-valence excited dication. In other words, the lifetime in a dimer is 
just an upper limit to what has to be expected in a bigger cluster. Therefore, in general the ICD lifetime 
is so short that alternative relaxation mechanisms are likely to be irrelevant. 

Our findings should have consequences for the interpretation of existing experimental data on 
core-hole relaxation in weakly bound clusters. Even more important might be the impact on the
fragmentation pattern of clusters, and larger molecules, exposed to the intense radiation of a 
future X-ray laser. If the laser pulse is to be used for structure determination through X-ray 
scattering, then the fragmentation timescale should be larger than the pulse length. Otherwise, 
the structure being measured is just an average over several Coulomb explosion stages.
Simulations, incorporating the usual photoionization and Auger processes, have shown that single 
protein molecules interacting with an intense femtosecond X-ray laser pulse disintegrate completely 
\cite{NeWo00}. Electron correlation effects like ICD lead to even higher charge states and, thus, to 
shorter fragmentation timescales. This will have to be taken into account in future experiments.

\acknowledgments
Financial support by the Deutsche Forschungsgemeinschaft is gratefully acknowledged.

\end{document}